\documentclass[letter,scriptaddress,twocolumn,showkeys]{revtex4}

    \usepackage{graphicx}
	\usepackage{amsmath}
	\usepackage{makeidx}
	\usepackage{amsfonts}
	\usepackage[ansinew]{inputenc}
	\usepackage[usenames,dvipsnames]{pstricks}
	\usepackage{subfigure}
	\usepackage{epsfig}
	\usepackage{pst-grad} 
	\usepackage{pst-plot} 
	\usepackage[colorlinks,hyperindex]{hyperref}
	\hypersetup
	{
		colorlinks,%
		citecolor=black,%
		linkcolor=black,%
		urlcolor=black,%
	}

\makeindex

\begin{document}

\title{Experimental observation of polarization-dependent optical vortex beams}

\author{S. Srisuphaphon}
\affiliation{Department of Physics, Faculty of Science, Burapha University, ChonBuri Province, 20131, Thailand}

\author{P. Panthong}
\affiliation{Department of Physics, Faculty of Science, Kasetsart University, Bangkok Province, 10900, Thailand}

\author{T. Photia}
\affiliation{Department of Physics, Faculty of Science, Kasetsart University, Bangkok Province, 10900, Thailand}

\author{W. Temnuch}
\affiliation{Department of Physics, Faculty of Science, Kasetsart University, Bangkok Province, 10900, Thailand}

\author{S. Chiangga}
\affiliation{Department of Physics, Faculty of Science, Kasetsart University, Bangkok Province, 10900, Thailand}

\author{S. Deachapunya}
\email{sarayut@buu.ac.th}
\affiliation{Department of Physics, Faculty of Science, Burapha University, ChonBuri Province, 20131, Thailand}

\date{today}

\begin{abstract}
We report the experimental demonstration of the induced polarization-dependent optical vortex beams. We use the Talbot configuration as a method to probe this effect. In particular, our simple experiment shows the direct measurement of this observation. Our experiment can exhibit clearly the combination between the polarization and orbital angular momentum (OAM) states of light. This implementation might be useful for further studies in the quantum system or quantum information.
\end{abstract}


\maketitle

Optical vortex or orbital angular momentum (OAM) of light has been widely used in optics~\cite{Allen1992,He1995,Grier2003,Zhan2009,Krenn2014}. Among various applications, the association between the polarization and OAM of light is promising. It can increase the degrees of freedom of optical states when working in field of quantum information. The spin-orbit non-separability or entanglement was performed~\cite{Souza2007,Borges2010,Toeppel2014,Aiello2015}. The systematic superposition of two vector Laguerre-Gaussian beams was presented with inhomogeneous polarization distribution~\cite{Vyas2013}. Also, non-coaxial superposition of vector vortex beams was conducted~\cite{Aadhi2016}. To make higher-order Poincar\'e sphere, one can add the polarization in terms of optical spin angular momentum to the total optical angular momentum that includes higher dimensional orbital angular momentum~\cite{Milione2011,Holleczek2011}. The study of polarization pattern of vector vortex beams with different topological charges was reported~\cite{Cardano2012}. The spin-to-orbital angular momentum conversion can be produced in ultra-thin metasurfaces~\cite{Bouchard2014}. The dynamics of optical spin-to-orbital angular momentum conversion were studied in uniaxial crystals~\cite{Brasselet2009}. Recently, a propagation model of vector beams generated by metasurfaces has been reported with both the Fresnel diffraction region and the Fraunhofer diffraction zone~\cite{Shu2016}. Also, the measurement of transverse force on a nano-cantilever exhibits another polarization-dependent contribution with the optical momentum~\cite{Antognozzi2016}, and the transverse spin density of light can be tuned by the mode and polarization state of the incoming beam~\cite{Neugebauer2015}.

Talbot effect~\cite{Talbot1836}, which is used in our work, is an optical near-field effect. The effect can be observed when a diffraction grating is illuminated by a coherent and monochromatic source such as a laser source. The grating self-image is appeared behind the grating at multiples of the specific distance, so-called the Talbot length ($L_T$) as well as the fractional Talbot effect can be occurred when the distance is the rational multiples of Talbot length, $iL_{T}/j$, where $i$ and $j$ are the integer number~\cite{Case2009,Srisuphaphon2015}. The Talbot effect is one of the crucial diffraction phenomena and has a lot of applications~\cite{Wen2013}.

\begin{figure}[htbp]
\centering
\includegraphics[width=8cm]{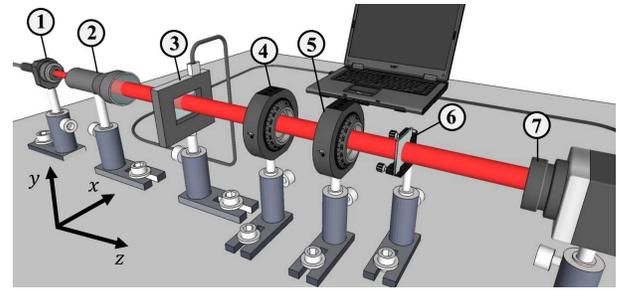}
\caption{Experimental setup. (1) laser diode; (2) beam expander; (3) spatial light modulator (SLM); (4) linear polarizer; (5) half-wave plate; (6) diffraction grating; and (7) Charge-coupled device camera (CCD).}
\label{Fig1}
\end{figure}

In this work, we introduce a simple method to combine the polarization and OAM states of light. The induced polarization-dependent optical vortex beam is performed by a typical polarizer and a half-wave plate, and using the Talbot effect as a detector. The OAM states are produced by a liquid crystal spatial light modulator (SLM). We demonstrate our idea with both positive and negative topological charges $l=1$, and $l=-1$. Our idea is simple and performed with small amount of optical elements.

\begin{figure}[htbp]
\centering
\includegraphics[width=8cm]{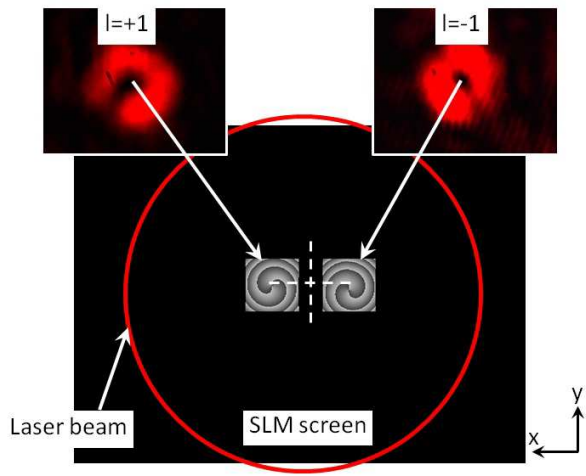}
\caption{Illustration of the key of our method. The gray-level images which contain both the vortex of $l=1$ (left), and $l=-1$ (right) are displayed on the SLM. This makes both optical vortices not aligned at the center of the Gaussian beam of laser. Then, the vortices become asymmetric shapes (the top insets). The dashed cross marks the center of the SLM screen. The optical vortices are formed at the distance of 170 cm behind the SLM and subsequently focused at 480 cm.}
\label{Fig2}
\end{figure}

\begin{figure*}[htb]
\centering
\includegraphics[width=15cm]{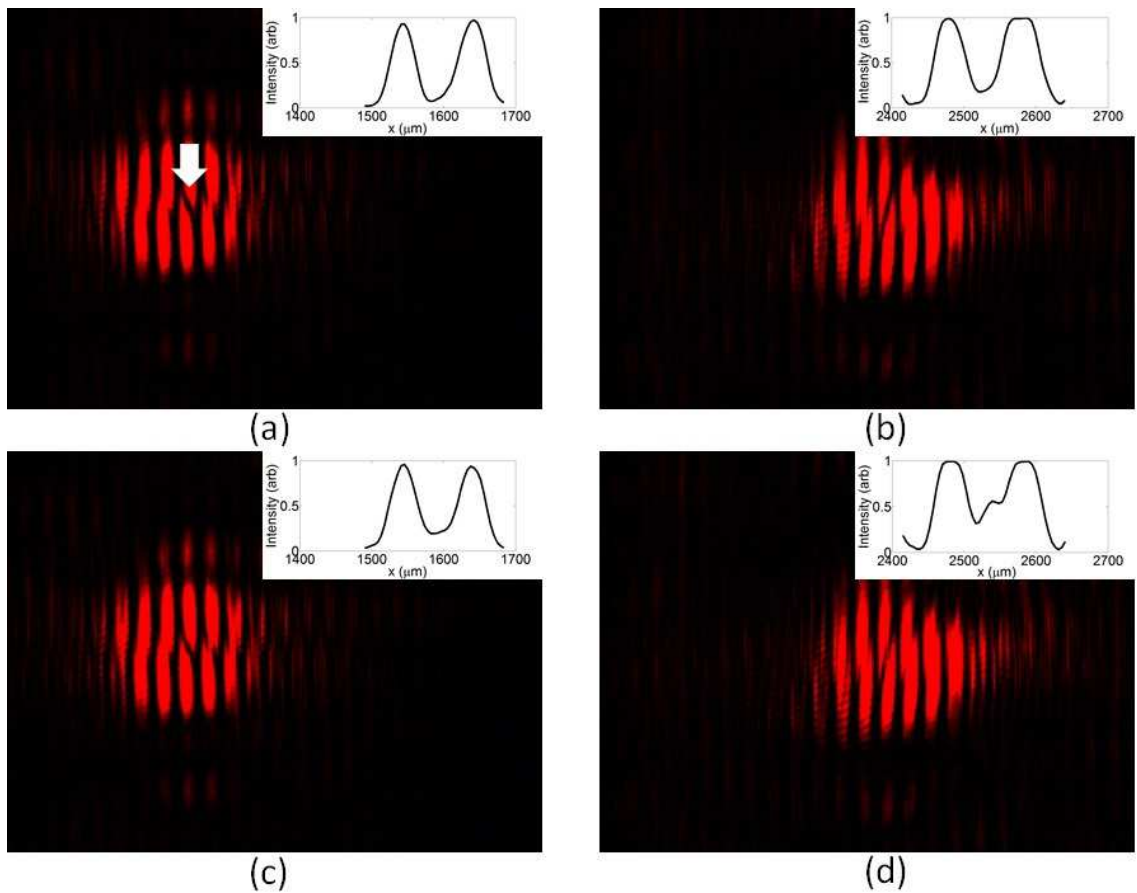}
\caption{The experimental Talbot patterns at $z=L_T=$63.5 mm, $d$=200 $\mu$m, and $\lambda$=630 nm (detail see text): $l=1$, with vertical polarization (a), $l=-1$, with vertical polarization (b), $l=1$, with horizontal polarization (c), $l=-1$, with horizontal polarization (d). The arrow points an example of the center dark stripe, which is used to determine in our work. The insets show the cross section of the center dark stripe.}
\label{Fig3}
\end{figure*}

We prove our idea with the experimental setup shown in Fig.~\ref{Fig1}. A diode laser ($\lambda$=630 nm) is used as a coherent source, which is then expanded and collimated by a beam expander, produced a Gaussian beam diameter of about 37 mm. A spatial light modulator (LC2012, Holoeye Photonics AG with resolution of 1024$\times$768 pixels, the pixel size of 36 $\mu$m) is used to produce an optical vortex by loading the SLM screen with a series of gray-level images (256 grey levels)~\cite{Panthong2016}. The laser beam is aligned to the center of SLM. Behind the SLM, a linear polarizer (LPNIR050-MP, Thorlabs) is used to indicate an initial polarization state. Subsequently, a half-wave plate (WPH10M-633, Thorlabs) can change the linear polarization state for study the polarization dependence of the OAM of light. A diffraction grating (chromium on glass, Edmund Optics inc., grating period ($d=200\mu m$), and opening fraction ($f=0.5$)) is placed at about $z=$300 cm behind the SLM in order to obtain the vortex of about 1.5 mm in diameter. The grating lines are aligned along the vertical y axis. The effect can be revealed only in the case of an asymmetric vortex and this is the key of our method. To make this asymmetric vortices, the gray-level images are not aligned to the center of the Gaussian beam of light or the SLM, as depicted in (Fig.~\ref{Fig2}). The vortex shapes are improved from our previous work~\cite{Panthong2016} by using the SLM mode with the proper focus. The Talbot length~\cite{Case2009} of $z=L_T=d^2/\lambda=$63.5 mm is set and a CCD (DCU223C, Thorlabs) is used to record the modified interference patterns (Fig.~\ref{Fig3}). The alignment for all optical elements has to be carefully concerned and the two methods can be used. First, all optical elements must have parallel faces and the back reflection of light is used. Second, the polarization-dependent observation to the OAM of light here is disappeared when the vortex beam is aligned to the center of the Gaussian beam of laser. In this case, the modified Talbot patterns remain unchanged with different polarization states.

We use the Talbot effect to identify the OAM of light~\cite{Panthong2016}. The dark (tilt) stripe, indicated by the arrow as shown in Fig.~\ref{Fig3} (a) for example, is used to define the OAM number of the vortex. The number of dark stripe provides the OAM number, while the direction presents the topological charge~\cite{Panthong2016}. For positive OAM, the upper parts of the interference pattern are shifted to the right and the lower parts are shifted oppositely. Fig.~\ref{Fig3} presents the experimental results with the vertical polarization (y axis) and the vortex charge of $l=1$ (a), compared to the horizontal polarization and the vortex charge of $l=1$ (c). The dark stripe becomes smaller and closer with the horizontal polarization, which is perpendicular to the grating lines. This can be explained as the influence of the spin or polarization can modify the OAM state. We confirm our assumption with the negative OAM number $l=-1$ for the vertical polarization, and the horizontal polarization as shown in Fig.~\ref{Fig3} (b), and (d), respectively. Both positive and negative charges present the similar trend. The insets show the cross section of the center dark stripe. The intensity at the center of the dark stripes is increased when the polarization state is perpendicular to the grating lines, i.e. the horizontal polarization in our case. The results, as shown here, can confirm our idea that the OAM of light can be controlled with the variation of an input polarization. In contrast to previous work~\cite{Lu2006}, we also show that the combination between the polarization and the OAM of light can produce the polarization-dependent Talbot effect with a low-density grating.

In conclusion, we can observe the relation between the polarization and the orbital angular momentum of light. The Talbot effect is used to probe this observation. Our measurement is obviously seen by a simple setup and permits higher degrees of freedom in optical experiments or even optical communications. In other words, our present proof-of-principle demonstrates spin-orbit coupling in optical systems. We also show that our technique can control the coupling. The coupling effect can be disappeared when the vortex beam is aligned to the center of the Gaussian beam of laser. Future experiments shall explore this observation in quantum experiments, such as single photon experiments~\cite{Song2011,Deachapunya2016}. The combination of high throughput beam and high efficiency single photon detection is needed for high-contrast interference patterns.

\begin{acknowledgements}
 S.D. acknowledges the support grant from the office of the higher education commission, the Thailand research fund (TRF), and Faculty of Science, Burapha university under contract number MRG5380264.
\end{acknowledgements}


\end{document}